\documentclass[12pt,prd,showpacs,tightenlines,nofootinbib]{revtex4}
\usepackage{amsfonts}
\usepackage{bm}
\usepackage{graphics}
\usepackage{amsmath}
\usepackage{epsfig}
\usepackage{bm}
\usepackage{multirow,booktabs}
\usepackage{slashed}

\begin{document}

\title{$\gamma\gamma\rightarrow M^{+}M^{-}(M=\pi, K)$ processes with twist-3 corrections in QCD}
\author{Cong Wang$^{1,2}$}
\author{Ming-Zhen Zhou$^{1}$\footnote{Corresponding Author}}
\email{zhoumz@swu.edu.cn}
\author{Hong Chen$^{1}$}
\affiliation{$^1$School of Physical Science and Technology, Southwest
University, Chongqing 400715,People's Republic of China\\
$^2$State Key Laboratory of Theoretical Physics, Institute of Theoretical Physics, Chinese Academy of Sciences, Beijing 100190, China}

\begin{abstract}
We study the $\gamma\gamma\rightarrow M^{+}M^{-}(M=\pi, K)$ processes with the contributions from the two-particle twist-2 and twist-3 distribution amplitudes of pion and kaon mesons on BHL prescription in the standard hard-scattering approach. The results show that the contributions from twist-3 parts are actually not power suppressed comparing with the leading-twist contributions in the low energy region and the cross sections agree well with the experimental data in the two-photon center-of-mass energy $W>2.8$ GeV. We also predict the cross section ratio $\sigma_{0}(\pi^{+}\pi^{-})/\sigma_{0}(K^{+}K^{-})$, which is compatible with the experimental data from TPC and Belle.
\end{abstract}

\pacs{12.38.Bx, 13.60.Le, 13.66.Bc}

\maketitle

\section{Introduction}

The common method of calculationing hard exclusive processes in Quantum Chromodynamics (QCD) was developed in Refs. \cite{Chernyak1,Chernyak2,Chernyak,Lepage,Brodsky2}. Especially, the exclusive processes at large momentum transfer have aroused much attention \cite{Chernyak,Lepage} in the last few years. As pioneers of the theoretical physicists, Brodsky and Lepage \cite{Lepage,Brodsky2} put forward a systematic analysis, involving angular dependence, helicity structure, normalization of elastic and inelastic form factors and large angle exclusive scattering amplitudes for hadrons and photons.

It is well known that the exclusive processes at large momentum transfer can afford definitely theoretical test for perturbative QCD. The two-photon processes, such as $\gamma^{\ast}\gamma \rightarrow$ hadrons and $\gamma\gamma \rightarrow$ hadron pairs at large momentum transfer, have attracted much attention in theoretical \cite{Brodsky2,Benayoun,Brodsky1,Chernyak4,Chernyak7} and experimental \cite{Aihara,Abe,Nakazawa,Nakazawa1,Heister,Mori} fields over the past few decades. Due to the pointlike structure of the photon, initial states are simple and controllable, and the strong interactions are presented only in final states. Such structure not only has an important role for understanding the nonperturbative construction of QCD, but also brings convenience to analysis of these exclusive hard scattering amplitudes and perturbative mechanisms.

In this work, we focus on the two-photon annihilation into pseudoscalar mesons $\gamma\gamma \rightarrow M^{+}M^{-}$. However, what's troubling theorists is that the cross sections predicted in theory are noticeably below the experimental data \cite{Nakazawa}. Brodsky and his collaborator provided the predictable results \cite{Brodsky2} as a $\sin^{-4}\theta$ dependence of the differential cross section and a $W^{-6}$ dependence of the total cross section. The similar theoretical predictions have been put forward in Chernyak's series of works on the two-photon exclusive processes \cite{Chernyak4,Chernyak7}. In 1986, Bene Ni$\check{\textrm{z}}$i$\acute{\textrm{c}}$ \cite{Nizic} was the first researcher who calculated the one-loop corrections for the two-photon exclusive channels at large momentum transfer, then Goran Duplan$\check{\textrm{c}}$i$\acute{\textrm{c}}$ \cite{Duplancic} perfected the leading-twist next-to-leading-order(NLO) radiative corrections. Their calculations indicated that the NLO corrections slightly change the leading-order predictions.

The early experimental work \cite{Aihara} has been suggested to test a QCD calculation in the model proposed by Brodsky and Lepage. Increasing interests for this problem, more experimental groups, for instance, TPC \cite{Aihara}, ALEPH \cite{Heister}, Belle \cite{Abe,Nakazawa,Nakazawa1} and so on have been attracted  to testify the results of theoretical analysis. Especially, the Belle collaboration systematically measured the two-photon collisions at the center-of-mass energy $2.4\ \textrm{GeV}<W<4.1\ \textrm{GeV}$ and the scattering-angle region $|\cos\theta|<0.6$ \cite{Nakazawa}. With the improvement of experimental accuracy, we hope to have a further understanding of these processes at high energy region and large scattering angle.

Besides the QCD radiation correction \cite{Nizic,Duplancic}, which is very minor in this process, the next-leading-order correction can also come from high Fock states or high twist distribution amplitudes of the hadron, and the later is considered in our work. From the naive point of view, the contributions of high twist distribution amplitudes are suppressed by the factor $1/Q^{2}$ for exclusive processes with large momentum transfer $Q$, but that has not always been the case. For example, the contributions from twist-3 distribution amplitudes are comparable with or even larger than the one from leading-twist distribution amplitude of light pseudoscalar meson in the $\chi_{cJ}\rightarrow \pi^{+}\pi^{-}, K^{+}K^{-}$ decay channels \cite{Zhou} and the pion/kaon electromagnetic form factor \cite{Wu1,Wu2}. More discussions for the high twist corrections of vector mesons are presented in Refs. \cite{Qiao,Pire1,Pire2}. In this work, the main experimental data of $\gamma\gamma \rightarrow M^{+}M^{-}$ processes come from the center-of-mass energy $W$ below $4.0\ \textrm{GeV}$ and the momentum transfer is not large enough. So it is necessary to investigate the contributions from the two-particle high twist distribution amplitudes for those channels. Our results also indicate that the twist-3 distribution amplitudes of pion and kaon make the significants contributions at the low energy region $W<6$ GeV. In the hard-scattering approach, there still exists the problem of the end-point singularity, which comes from twist-3 $\varphi_{p}(x)$ term of pseudoscalar mesons \cite{HTao2}, and we take into account the meson distribution amplitudes with BHL prescription where the distribution amplitudes are rewritten with the exponential suppression factors to avoid the end-point effect.

The structure of this paper is arranged as follows. In Section \uppercase\expandafter{\romannumeral2}, we present our calculation of the hard-scattering amplitude at tree level. A brief model of the twist-2 and twist-3 distribution amplitudes with BHL scheme is presented in Section \uppercase\expandafter{\romannumeral3}. The Section \uppercase\expandafter{\romannumeral4} is the numerical analysis and the last Section is the conclusion to this work. The invariant amplitudes for $\gamma\gamma\rightarrow \pi^{+}\pi^{-}, K^{+}K^{-}$ are given in Appendix A.

\section{calculation of hard-scattering amplitudes}

\begin{figure}
\centering
     \includegraphics[width=0.35\textwidth, height=0.27\textwidth]{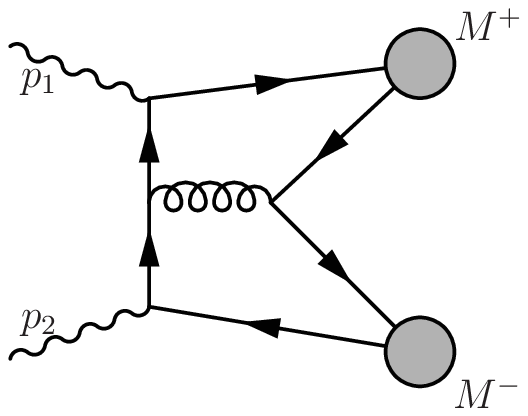}
     \includegraphics[width=0.35\textwidth, height=0.27\textwidth]{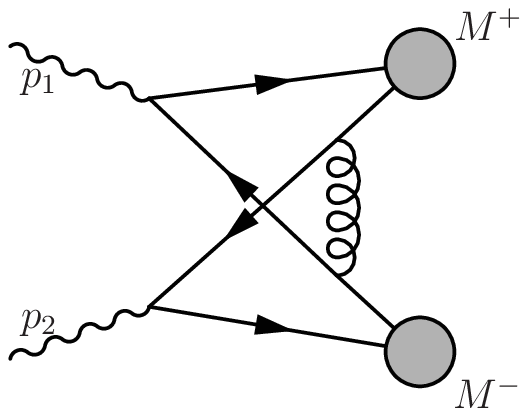}
\caption{Feynman diagrams for hard QCD contributions to $\gamma\gamma\rightarrow M^{+}M^{-}$.}
\label{feyndiag}
\end{figure}

In this section, we want to recalculate the hard-scattering amplitudes for
\begin{eqnarray}\label{1.1}
\gamma_{1}(p_{1},\varepsilon^{\lambda_{1}}_{1})+\gamma_{2}(p_{2},\varepsilon^{\lambda_{2}}_{2}) \rightarrow M^{+}(p_{3})+M^{-}(p_{4}), \quad (M=\pi,K)
\end{eqnarray}
where the initial photons are real with the different polarized vectors and the final states are flavor-nonsinglet helicity-zero mesons. $p_{1},p_{2}$ and $p_{3},p_{4}$ are four-momenta for the initial photons and the final pseudoscalar mesons, respectively. $\varepsilon^{\lambda_{1}}_{1}(\varepsilon^{\lambda_{2}}_{2})$ is the polarized vector of the photon and the scripts $\lambda_{1}(\lambda_{2})=\pm1$ represent two transversal photons. This process is described by the $\gamma\gamma \rightarrow (q\overline{q})+(q\overline{q})$ amplitude which is illustrated by two typical lowest order Feynman diagrams in Fig. \ref{feyndiag}. The left typical diagram has twelve topological structures and the right typical diagram has eight topological structures by various symmetries from photons, gluons and quarks exchanging.

In the two-photon center-of-mass frame, we choose the direction of two-photon collision as Z-axis with the total energy denoted by $W$ and the scattering angle denoted by $\theta$. And then, the four-momenta of incoming and outgoing particles are given as follows
\begin{eqnarray}\label{1.2}
p_{1}=\frac{W}{2}\left(1,0,0,1\right),  &&  p_{2}=\frac{W}{2}\left(1,0,0,-1\right),\nonumber\\
\quad p_{3}=\frac{W}{2}\left(1,\sin\theta,0,\cos\theta\right), && p_{4}=\frac{W}{2}\left(1,-\sin\theta,0,-\cos\theta\right),
\end{eqnarray}
where the masses of pion and kaon are canceled in the chiral limit. The polarization states of photons are taken to be
\begin{eqnarray}\label{1.3}
\varepsilon_{1}^{+}(p_{1})=\frac{1}{\sqrt{2}}(0,-1,-i,0),&& \quad \varepsilon_{1}^{-}(p_{1})=\frac{1}{\sqrt{2}}(0,1,-i,0),\nonumber\\
\varepsilon_{2}^{+}(p_{2})=\frac{1}{\sqrt{2}}(0,1,-i,0),&& \quad \varepsilon_{2}^{-}(p_{2})=\frac{1}{\sqrt{2}}(0,-1,-i,0).
\end{eqnarray}

With the above choices, we work out the expression of cross section with the two-particle twist-3 distribution amplitudes of the final pion mesons in the hard-scattering approach and it is written as
\begin{eqnarray}\label{1.4}
\frac{d\sigma(\gamma\gamma \rightarrow \pi^{+}\pi^{-})}{d \cos\theta}=\frac{1}{32\pi W^{2}}\frac{1}{4}\sum_{\lambda_{1},\lambda_{2}=\pm1} \left|A_{\lambda_{1}\lambda_{2}}\right|^{2},
\end{eqnarray}
\begin{eqnarray}\label{1.5}
A_{\lambda_{1}\lambda_{2}}(W,\theta)=\int_{0}^{1}\int_{0}^{1}dxdy\sum_{ij}\phi^{i}_{\pi}(x,\mu_{F}^2)
T^{ij}_{\lambda_{1}\lambda_{2}}(x,y,W,\theta)\phi^{j}_{\pi}(y,\mu_{F}^2),
\end{eqnarray}
where $A_{\lambda_{1}\lambda_{2}}$ is the transitional matrix element of the two-photon scattering process. $\phi^{i}_{\pi}(x,\mu_{F}^{2})$ is the pion meson distribution amplitude with the longitudinal momentum fraction $x$ and the factorization scale $\mu_{F}$. The script $i=\pi$ represents the twist-2 distribution amplitude and the scripts $i=p,\sigma$ represent two twist-3 distribution amplitudes. The function $T^{ij}_{\lambda_{1}\lambda_{2}}(x,y,W,\theta)$ is the invariant amplitude from the different distribution amplitudes with the scripts $i$ and $j$ .

To calculate the transitional matrix element, we take the vacuum saturation approximation and use the Feirz identity
\begin{eqnarray}\label{1.6}
\overline{q}_{1\alpha}q_{2\beta}&=&\frac{1}{4}I_{\beta\alpha}(\overline{q}_{1}q_{2})-\frac{1}{4}(i\gamma_{5})_{\beta\alpha}(\overline{q}_{1}i\gamma^{5}q_{2})
+\frac{1}{4}(\gamma_{\mu})_{\beta\alpha}(\overline{q}_{1}\gamma^{\mu}q_{2})\nonumber\\
&-&\frac{1}{4}(\gamma_{\mu}\gamma_{5})_{\beta\alpha}(\overline{q}_{1}\gamma^{\mu}\gamma^{5}q_{2})
+\frac{1}{8}(\sigma_{\mu\nu}\gamma_{5})_{\beta\alpha}(\overline{q}_{1}\sigma^{\mu\nu}\gamma^{5}q_{2}),
\end{eqnarray}
where $q_{1}$ and $q_{2}$ are the quark fields. There are only three terms with the matrix $\gamma_5$ that have contributions for scattering amplitudes in the  $\gamma\gamma\rightarrow \pi^{+}\pi^{-}$ process by the parity analysis. Finally, we find that the scripts of nonzero terms are $ij=\, \pi\pi, \, pp, \, p\sigma, \, \sigma p, \, \sigma\sigma$ for the invariant amplitude  $T^{ij}_{\lambda_{1}\lambda_{2}}(x,y,W,\theta)$ of this reaction. The twist-2 and twist-3 distribution amplitudes of pion are defined as the following relations
\begin{eqnarray}\label{1.7}
\langle \pi^{+}(p)|\overline{u}(z_{1})\gamma^{\mu}\gamma^{5}d(z_{2})|0\rangle=-if_{\pi}p^{\mu}\int_{0}^{1}dx\ e^{i(x\ p\cdot z_{1}+(1-x)\ p\cdot z_{2})} \phi_{\pi}^{\pi}(x),
\end{eqnarray}
\begin{eqnarray}\label{1.8}
\langle \pi^{+}(p)|\overline{u}(z_{1})i\gamma^{5}d(z_{2})|0\rangle=f_{\pi}\mu_{\pi}\int_{0}^{1}dx\ e^{i(x\ p\cdot z_{1}+(1-x)\ p\cdot z_{2})} \phi_{\pi}^{p}(x),
\end{eqnarray}
\begin{eqnarray}\label{1.9}
\langle \pi^{+}(p)|\overline{u}(z_{1})\sigma^{\mu\nu}\gamma^{5}d(z_{2})|0\rangle
&=&\frac{i f_{\pi}\mu_{\pi}}{6}[p^{\mu}(z_{1}-z_{2})^{\nu}-p^{\nu}(z_{1}-z_{2})^{\mu}]\nonumber\\
& & \times\int_{0}^{1}dx\ e^{i(x\ p\cdot z_{1}+(1-x)\ p\cdot z_{2})} \phi_{\pi}^{\sigma}(x),
\end{eqnarray}
by the expanded form of hadronic matrix element \cite{Branu1,Branu2,Nagashima,Beneke}, where $f_{\pi}$ is the decay constant of pion, the parameter $\mu_{\pi}=\frac{m_{\pi}^{2}}{m_{u}+m_{\overline{d}}}$ is proportional to the chiral condensate and the variable $x$ is the meson momentum fraction.

With the definitions of distribution amplitudes for final state mesons, the transitional matrix element of the $\gamma\gamma\rightarrow \pi^{+}\pi^{-}$ process is calculated and the invariant amplitude $T^{ij}_{\lambda_{1}\lambda_{2}}(x,y,W,\theta)$ is represented as
\begin{eqnarray}\label{1.10}
T^{ij}_{\lambda_{1}\lambda_{2}}(x,y,W,\theta)=\frac{16}{3}\pi^{2}\alpha\alpha_{s}(\mu_{R}^{2})C_{F}\widehat{T}^{ij}_{\lambda_{1}\lambda_{2}}
\frac{1}{l_{1}^{2}l_{2}^{2}l_{3}^{2}},
\end{eqnarray}
where $C_{F}=\frac{4}{3}$ is the color factor and the masses of light quarks are canceled. $\alpha$ is the electromagnetic coupling constant and $\alpha_{s}(\mu_{R}^{2})$ is the strong coupling constant with the renormalization scale $\mu_{R}$.  The operator $\widehat{T}^{ij}_{\lambda_{1}\lambda_{2}}$ is related to the two-pion materialization of two photons with different twist distribution amplitudes and has the diverse expression from the various Feynman diagram. For instance, it is given as
\begin{eqnarray}\label{1.11}
\widehat{T}^{\pi\pi}_{\lambda_{1}\lambda_{2}}=\frac{1}{16}f_{\pi}^{2}\mathrm{Tr}[\slashed{\varepsilon}_{1}\slashed{l}_{1}\gamma_{\rho}\slashed{l}_{2}
\slashed{\varepsilon}_{2}\gamma_{5}\slashed{p}_{4}\gamma^{\rho}\gamma_{5}\slashed{p}_{3}]\nonumber,
\end{eqnarray}
\begin{eqnarray}\label{1.12}
\widehat{T}^{pp}_{\lambda_{1}\lambda_{2}}=\frac{1}{16}f_{\pi}^{2}\mu_{\pi}^{2}\mathrm{Tr}[\slashed{\varepsilon}_{1}\slashed{l}_{1}\gamma_{\rho}\slashed{l}_{2}
\slashed{\varepsilon}_{2}\gamma_{5}\gamma^{\rho}\gamma_{5}]\nonumber,
\end{eqnarray}
\begin{eqnarray}\label{1.13}
\widehat{T}^{p\sigma}_{\lambda_{1}\lambda_{2}}=-\frac{1}{96}f_{\pi}^{2}\mu_{\pi}^{2}
p_{3}^{\alpha}(\frac{\partial}{\partial l_{1\beta}}-\frac{\partial}{\partial l_{3\beta}})
\mathrm{Tr}[\slashed{\varepsilon}_{1}\slashed{l}_{1}\gamma_{\rho}\slashed{l}_{2}
\slashed{\varepsilon}_{2}\gamma_{5}\gamma^{\rho}\gamma_{5}\sigma_{\alpha\beta}]\nonumber,
\end{eqnarray}
\begin{eqnarray}\label{1.14}
\widehat{T}^{\sigma p}_{\lambda_{1}\lambda_{2}}=-\frac{1}{96}f_{\pi}^{2}\mu_{\pi}^{2}
p_{4}^{\mu}(\frac{\partial}{\partial l_{3\nu}}-\frac{\partial}{\partial l_{2\nu}})
\mathrm{Tr}[\slashed{\varepsilon}_{1}\slashed{l}_{1}\gamma_{\rho}\slashed{l}_{2}
\slashed{\varepsilon}_{2}\gamma_{5}\sigma_{\mu\nu}\gamma^{\rho}\gamma_{5}]\nonumber,
\end{eqnarray}
\begin{eqnarray}\label{1.15}
\widehat{T}^{\sigma\sigma}_{\lambda_{1}\lambda_{2}}&=&\frac{1}{576}f_{\pi}^{2}\mu_{\pi}^{2}
p_{3}^{\alpha}(\frac{\partial}{\partial l_{1\beta}}-\frac{\partial}{\partial l_{3\beta}})p_{4}^{\mu}(\frac{\partial}{\partial l_{3\nu}}-\frac{\partial}{\partial l_{2\nu}})
\mathrm{Tr}[\slashed{\varepsilon}_{1}\slashed{l}_{1}\gamma_{\rho}\slashed{l}_{2}
\slashed{\varepsilon}_{2}\gamma_{5}\sigma_{\alpha\beta}\gamma^{\rho}\gamma_{5}\sigma_{\alpha\beta}],
\end{eqnarray}
in the left diagram of Fig. \ref{feyndiag}, where $\varepsilon_{1}$ and $\varepsilon_{2}$ are the abbreviations of polarized vectors $\varepsilon_{1}^{\lambda_{1}}(p_{1})$ and $\varepsilon_{2}^{\lambda_{2}}(p_{2})$ of initial photons, respectively. The partial $(\frac{\partial}{\partial l_{m\nu}}-\frac{\partial}{\partial l_{n\nu}})$ ($m,n=1,2,3$) comes  from the $(z_{1}-z_{2})^{\nu}$ of Eq. (\ref{1.9}). The momenta of quark propagators are represented by $l_{1}$, $l_{2}$ and the gluon propagator is represented by $l_{3}$. In this diagram, they are written as $l_{1}=-p_{1}+(1-x)p_{3}$, $l_{2}=p_{2}-y p_{4}$ and $l_{3}=-x p_{3}-(1-y)p_{4}$. Finally, we obtain the invariant amplitudes $T^{ij}_{\lambda_{1}\lambda_{2}}(x,y,W,\theta)$ with the scripts $ij=\, \pi\pi, \, pp, \, p\sigma, \, \sigma p, \, \sigma\sigma$ by the sum of twenty Feynman diagrams and their detailed expressions are shown in Appendix A.

\section{the distribution amplitudes of pseudoscalar mesons}

The twist-2 and twist-3 distribution amplitudes of pseudoscalar mesons are taken as the main nonperturbative input parameters in the above calculations of hard scattering amplitude. In this section, we give a brief model for them in BHL scheme \cite{Huang1}. Equating the off-shell propagator to connect the equal-time wave function in the rest frame and the light-cone wave function in the infinite momentum frame, the wave function for quark-antiquark systems is obtained
\begin{eqnarray}\label{2.1}
\Psi(x,\mathbf{k}_{\bot})\propto\text{exp}\left[-\frac{1}{8\beta^{2}}
\left(\frac{\mathbf{k}_{\bot}^{2}+m_{1}^{2}}{x}+\frac{\mathbf{k}_{\bot}^{2}+m_{2}^{2}}{1-x}\right)\right]
\end{eqnarray}
from the harmonic oscillator model at the rest frame, where $m_{i}$ is the constitute quark mass. The $\beta$ refers to the harmonic parameter which can be obtained from the definition of the average of quark transverse momentum squared
\begin{eqnarray}\label{2.2}
\langle\mathbf{k}_{\perp}^{2}\rangle_{M}=\frac{f_{M}^{2}}{24}\int dx\frac{d^{2}\mathbf{k}_{\perp}}{16\pi^{3}}|\mathbf{k}_{\perp}^{2}||\Psi_{M}(x,\mathbf{k}_{\perp})|^{2}/P^{M}_{q\overline{q}},
\end{eqnarray}
where $M=\pi$ for pion and $M=K$ for kaon mean the leading-twist wave functions $\Psi^{\pi}_{\pi}$ and $\Psi^{K}_{K}$. The decay constants are $f_{\pi}=0.132$\ GeV for pion and $f_{K}=0.160$\ GeV for kaon, respectively.
In the Ref. \cite{Huang1}, $\langle\mathbf{k}_{\perp}^{2}\rangle_{\pi}$ and $\langle\mathbf{k}_{\perp}^{2}\rangle_{K}$ are all given as $(0.356\
\textrm{GeV})^{2}$ approximately. The probability of finding the $q\overline{q}$ leading-twist Fock state in the pseudoscalar meson is not larger than unity
\begin{eqnarray}\label{2.3}
P_{q\overline{q}}^{M}=\frac{f_{M}^{2}}{24}\int dx\frac{d^{2}\mathbf{k}_{\perp}}{16\pi^{3}}|\Psi_{M}(x,\mathbf{k}_{\perp})|^{2}\leq1.
\end{eqnarray}

The classical forms of twist-2 and twist-3 wave functions with BHL prescription are widely considered and have an immediate advantage to solve the end-point singularity by the exponential suppression in $x=0$ and $x=1$ point. In our work, we take the twist-2 wave functions of pion and kaon with first three terms Gegenbauer polynomials and they are characterized as
\begin{eqnarray}\label{2.4}
\Psi_{\pi}^{\pi}(x,\mathbf{k}_{\perp})=A_{\pi}^{\pi}\left[1+B_{\pi}^{\pi}C_{2}^{\frac{3}{2}}(\xi)+C_{\pi}^{\pi}C_{4}^{\frac{3}{2}}(\xi)\right]
\text{exp}\left[-\frac{\mathbf{k}_{\perp}^{2}+m_{q}^{2}}{8\beta_{\pi}^{2}x(1-x)}\right],
\end{eqnarray}

\begin{eqnarray}\label{2.5}
\Psi_{K}^{K}(x,\mathbf{k}_{\perp})=A_{K}^{K}\left[1+B_{K}^{K}C_{1}^{\frac{3}{2}}(\xi)+C_{K}^{K}C_{2}^{\frac{3}{2}}(\xi)\right]
\text{exp}\left[-\frac{1}{8\beta_{K}^{2}}\left(\frac{\mathbf{k}_{\perp}^{2}+m_{q}^{2}}{x}+\frac{\mathbf{k}_{\perp}^{2}+m_{s}^{2}}{1-x}\right)\right],
\end{eqnarray}
where $C_{n}^{\frac{3}{2}}(\xi)$ is relevant to Gegenbauer polynomials with the relationship $\xi=2 x-1$ and we take $n=2,4$ in the pion case for the $SU(2)$ isotopic symmetry and $n=1,2$ in the kaon case for the $SU(3)$-flavor symmetry breaking. The constitute quark masses $m_{q}=0.30$ \textrm{GeV} $(q=u,d)$ and $m_{s}=0.45$ \textrm{GeV} \cite{Zhou} are given in the above formulas. To simplify the following numerical analysis, we write twist-3 wave functions as

\begin{eqnarray}\label{2.6}
\Psi_{\pi}^{p}(x,\mathbf{k}_{\perp})=\frac{A_{\pi}^{p}}{x(1-x)}\text{exp}\left[-\frac{\mathbf{k}_{\perp}^{2}+m_{q}^{2}}{8\beta_{\pi}^{2}x(1-x)}\right],
\end{eqnarray}

\begin{eqnarray}\label{2.7}
\Psi_{\pi}^{\sigma}(x,\mathbf{k}_{\perp})=A_{\pi}^{\sigma}\text{exp}\left[-\frac{\mathbf{k}_{\perp}^{2}+m_{q}^{2}}{8\beta_{\pi}^{2}x(1-x)}\right],
\end{eqnarray}
and

\begin{eqnarray}\label{2.8}
\Psi_{K}^{p}(x,\mathbf{k}_{\perp})=\frac{A_{K}^{p}}{x(1-x)}
\text{exp}\left[-\frac{1}{8\beta_{K}^{2}}\left(\frac{\mathbf{k}_{\perp}^{2}+m_{q}^{2}}{x}+\frac{\mathbf{k}_{\perp}^{2}+m_{s}^{2}}{1-x}\right)\right],
\end{eqnarray}

\begin{eqnarray}\label{2.9}
\Psi_{K}^{\sigma}(x,\mathbf{k}_{\perp})=A_{K}^{\sigma}
\text{exp}\left[-\frac{1}{8\beta_{K}^{2}}\left(\frac{\mathbf{k}_{\perp}^{2}+m_{q}^{2}}{x}+\frac{\mathbf{k}_{\perp}^{2}+m_{s}^{2}}{1-x}\right)\right],
\end{eqnarray}
for the pion and kaon, respectively.

The above wave functions of twist-2 and twist-3 for pion and kaon follow the normalized condition
\begin{eqnarray}\label{2.10}
\int dx\frac{d^{2}\mathbf{k}_{\perp}}{16\pi^{3}}\Psi_{M}^{i}(x,\mathbf{k}_{\perp})=1,
\end{eqnarray}
where the scripts $i=\pi(K),p,\sigma$ refer to different twist wave functions and the script $M=\pi(K)$ refers to the pion(kaon) meson. By integrating the transverse momentum of wave function, corresponding distribution amplitude is acquired from the relationship
\begin{eqnarray}\label{2.11}
\phi^{i}_{M}(x,\mu_{F}^2)=\int_{|\mathbf{k}_{\perp}|<\mu_{F}}\frac{d^{2}\mathbf{k}_{\perp}}{16\pi^{3}}\Psi^{i}_{M}(x,\mathbf{k}_{\perp}),
\end{eqnarray}
where $\mu_{F}$ is the upper limit of integral which refers to the ultraviolet cutoff.
Substituting Eq. (\ref{2.4})-(\ref{2.9}) into Eq. (\ref{2.11}), we obtain the meson distribution amplitudes
\begin{eqnarray}\label{2.12}
\phi_{\pi}^{\pi}(x)=\frac{A_{\pi}^{\pi}\beta_{\pi}^{2}}{2\pi^{2}}x(1-x)[1+B_{\pi}^{\pi}C_{2}^{\frac{3}{2}}(\xi)+C_{\pi}^{\pi}C_{4}^{\frac{3}{2}}(\xi)]
\text{exp}\left[-\frac{m_{q}^{2}}{8\beta_{\pi}^{2}x(1-x)}\right],
\end{eqnarray}
\begin{eqnarray}\label{2.13}
\phi_{\pi}^{p}(x)=\frac{A_{\pi}^{p}\beta_{\pi}^{2}}{2\pi^{2}}\text{exp}\left[-\frac{m_{q}^{2}}{8\beta_{\pi}^{2}x(1-x)}\right],
\end{eqnarray}
\begin{eqnarray}\label{2.14}
\phi_{\pi}^{\sigma}(x)=\frac{A_{\pi}^{\sigma}\beta_{\pi}^{2}}{2\pi^{2}}x(1-x)\text{exp}\left[-\frac{m_{q}^{2}}{8\beta_{\pi}^{2}x(1-x)}\right],
\end{eqnarray}
for pion and
\begin{eqnarray}\label{2.15}
\phi_{K}^{K}(x)=\frac{A_{K}^{K}\beta_{K}^{2}}{2\pi^{2}}x(1-x)[1+B_{K}^{K}C_{1}^{\frac{3}{2}}(\xi)+C_{K}^{K}C_{2}^{\frac{3}{2}}(\xi)]
\text{exp}\left[-\frac{(1-x)m_{q}^{2}+x m_{s}^{2}}{8\beta_{K}^{2}x(1-x)}\right],
\end{eqnarray}
\begin{eqnarray}\label{2.16}
\phi_{K}^{p}(x)=\frac{A_{K}^{p}\beta_{K}^{2}}{2\pi^{2}}\text{exp}\left[-\frac{(1-x)m_{q}^{2}+x m_{s}^{2}}{8\beta_{K}^{2}x(1-x)}\right],
\end{eqnarray}
\begin{eqnarray}\label{2.17}
\phi_{K}^{\sigma}(x)=\frac{A_{K}^{\sigma}\beta_{K}^{2}}{2\pi^{2}}x(1-x)\text{exp}\left[-\frac{(1-x)m_{q}^{2}+x m_{s}^{2}}{8\beta_{K}^{2}x(1-x)}\right],
\end{eqnarray}
for kaon, respectively. The above coefficients $A_{M}^{i},B_{M}^{i},C_{M}^{i},A_{M}^{p},A_{M}^{\sigma}$ and the harmonic parameters $\beta_{M}$ are worked out in the following discussion.

With the method of nonlocal operator product expansion and conformal symmetry, the distribution amplitudes of the pion \cite{Ball2} and kaon \cite{Ball3} have been studied and the general form of the leading twist distribution amplitude with the expansion of Gegenbauer polynomials were described as
\begin{eqnarray}\label{2.18}
\phi_{M}^{i}(x,\mu^{2}_{F})=6x(1-x)\left(1+\sum_{n=1}^{\infty}a_{n}^{i}(\mu^{2}_{F})C_{n}^{\frac{3}{2}}(2x-1)\right).
\end{eqnarray}
To leading logarithmic accuracy, the coefficients of nonperturbative Gegenbauer polynomials $a_{n}^{i}$ renormalize multiplicatively with
\begin{eqnarray}\label{2.19}
a_{n}^{i}(\mu^{2}_{F})=L^{\frac{\gamma_{n}^{(0)}}{\beta_{0}}}a_{n}^{i}(\mu_{0}^{2}),
\end{eqnarray}
where $L=\frac{\alpha_{s}(\mu^{2}_{F})}{\alpha_{s}(\mu_{0}^{2})}$, $\beta_{0}=\frac{11 N_{c}-2 N_{f}}{3}$ with the quark colour number $N_{c}=3$ and the quark flavor number $N_{f}=4$. The script n takes even numbers for the pion and positive integers for the kaon. The anomalous dimension $\gamma_{n}^{(0)}$ can be expressed as
\begin{eqnarray}\label{2.20}
\gamma_{n}^{(0)}=4C_{F}\left(\psi(n+2)+\gamma_{E}-\frac{3}{4}-\frac{1}{2(n+1)(n+2)}\right).
\end{eqnarray}
The moments of leading twist distribution amplitudes are defined as the following expression
\begin{eqnarray}\label{2.21}
\langle\xi^{n}\rangle^{i}_{M}=\frac{1}{2}\int_{-1}^{1}\xi^{n}\phi^{i}_{M}(\xi)d\xi
\end{eqnarray}
with $\xi=2x-1$. Those coefficients of Gegenbauer polynomials $a_{n}^{i}$ were mainly calculated by QCD sum rules\cite{Chernyak3,Ball2,Ball3,Huang3} and we take their values

\begin{eqnarray}\label{2.22}
a_{2}^{\pi}(1\ \textrm{GeV})=0.25\pm0.15,&& \quad a_{4}^{\pi}(1\ \textrm{GeV})=0.04\pm0.11,\nonumber\\
a_{1}^{K}(1\ \textrm{GeV})=0.06\pm0.03,&& \quad a_{2}^{K}(1\ \textrm{GeV})=0.25\pm0.15.
\end{eqnarray}
Establishing the equations by the definition of moments Eq.\ (\ref{2.21}) which combine the leading twist distribution amplitudes Eqs.\ (\ref{2.12}) and (\ref{2.15}) in our model and the naive distribution amplitudes Eq.\ (\ref{2.18}), we solve the harmonic parameters $\beta_{M}$ due to the average of quark transverse momentum squared Eq.\ (\ref{2.2}) and the coefficients $A_{M}^{i},B_{M}^{i},C_{M}^{i}$ of twist-2 distribution amplitudes for the pion and kaon from the BHL scheme with the help of the coefficients of Gegenbauer polynomials Eq.\ (\ref{2.22}). And then the coefficients $A_{M}^{p},A_{M}^{\sigma}$ of twist-3 distribution amplitudes can be obtained directly from the normalized condition Eq.\ (\ref{2.10}).

\begin{figure}
\centering
     \includegraphics[width=0.48\textwidth, height=0.36\textwidth]{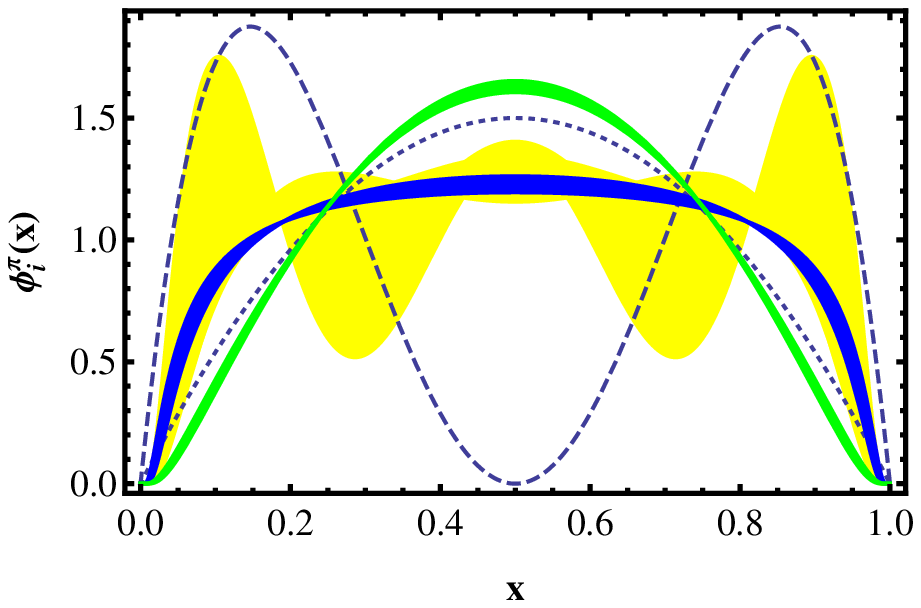}
     \includegraphics[width=0.48\textwidth, height=0.36\textwidth]{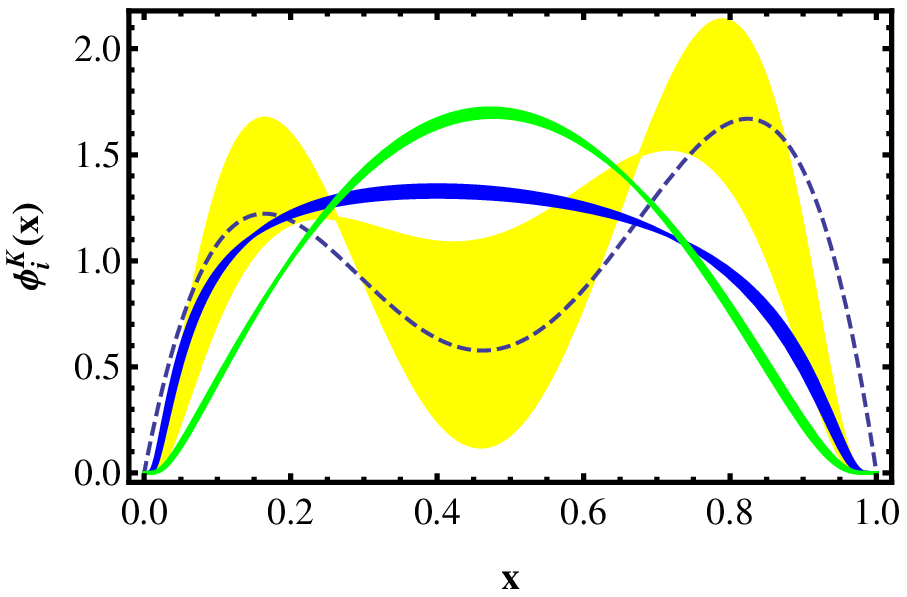}
\caption{The characteristic shapes of distribution amplitudes for pion and kaon in BHL frame. Left panel: Yellow band: $\phi_{\pi}^{\pi}$; Blue band: $\phi_{\pi}^{p}$; Green band: $\phi_{\pi}^{\sigma}$. Dashed line: the CZ distribution amplitude, $\phi_{\pi}^{CZ}(x)=30x_{d}x_{u}(x_{d}-x_{u})^{2}$, Dotted line:
the asymptotic distribution amplitude, $\phi_{\pi}^{asy}(x)=6x(1-x)$. Right panel: same for $\phi_{K}^{i}$; $\phi_{K}^{CZ}(x)=30x_{s}x_{u}[0.6(x_{s}-x_{u})^{2}+0.08+0.08(x_{s}-x_{u})]$.}
\label{DApik}
\end{figure}

Thinking about the Eq. (\ref{2.19}) in the analysis of distribution amplitudes, we actually get a series of parameters which change with the scale value $\mu^{2}_{F}=W^{2}$ and they will be used in the next numerical analysis to the $\gamma\gamma\rightarrow \pi^{+}\pi^{-}, K^{+}K^{-}$ processes. The coefficients of twist-2 and twist-3 distribution amplitudes involved in our calculations are too numerous to table them adequately and Fig. \ref{DApik} is illustrated with different distribution amplitudes in the energy $W\in(1,6)\ \textrm{GeV}$. As can be seen from Fig. \ref{DApik}, distribution amplitudes in BHL prescription are obviously different from the common forms in the point $x=0$ and $1$. The end-point singularities coming from the hard-scattering amplitudes are solved in our model. The banded graphics of distribution amplitudes are given out with the variable parameters: the strong coupling constant $\alpha_{s}$ with $\Lambda_{QCD}\in(0.15,0.3)\ \textrm{GeV}$, the center-of-mass energy $W\in(1,6)\ \textrm{GeV}$ and the Gegenbauer coefficients $a^{i}_{n}$ from Eq. (\ref{2.22}). Especially, the area of the pion $\phi_{\pi}^{\pi}$ is more complicated than one of the kaon $\phi_{K}^{K}$, since we take Gegenbauer polynomials to the $C_{4}^{\frac{3}{2}}$ term with the coefficient $a_{4}^{\pi}$ changing from positive value to negative value for the pion, but the coefficient $a_{2}^{K}$ is always the positive one for the kaon. The graphics of twist-3 distribution amplitudes without Gegenbauer polynomials are simple for the pion and kaon.

\section{numerical analysis}
To analyze the scattering cross section of two photon annihilation into pseudoscalar pairs numerically, we take the electromagnetic coupling constant $\alpha=\frac{1}{137}$ and the QCD running coupling constant calculated to two-loop accuracy is given by
\begin{eqnarray}\label{3.1}
\alpha_{s}(\mu_{R}^{2})=\frac{4\pi}{\beta_{0}\textrm{ln}(\frac{\mu_{R}^{2}}{\Lambda_{QCD}^{2}})}
\left\{1-\frac{\beta_{1}}{\beta_{0}^{2}}\frac{\textrm{lnln}(\frac{\mu_{R}^{2}}{\Lambda_{QCD}^{2}})}{\textrm{ln}(\frac{\mu_{R}^{2}}{\Lambda_{QCD}^{2}})}\right\}
\end{eqnarray}
with $\beta_{0}=\frac{11 N_{c}-2 N_{f}}{3}$ and $\beta_{1}=\frac{34 N_{c}^{3}-13 N_{c}^{2} N_{f}+3 N_{f}}{3 N_{c}}$. The QCD scale is taken as $\Lambda_{QCD}\in(0.15,0.3)$\ GeV and the interaction scale is chosen as $\mu_{R}^{2}=W^{2}$ in this work.

Next, we focus on the chiral enhancing scale $\mu_{M}$, which is an important parameter with the sensitively influence to the contribution from twist-3 parts of the pseudoscalar meson in the $\gamma\gamma\rightarrow M^{+}M^{-}$ process. According to its definition
\begin{eqnarray}\label{3.2}
\mu_{M}=\frac{m_{M}^{2}}{m_{q}+m_{\overline{q}}},\ (M=\pi,K; q,\overline{q}=u,d,s),
\end{eqnarray}
and one-loop expression for the running quark mass in the $\overline{\textrm{MS}}$ scheme \cite{Buras}
\begin{eqnarray}\label{3.3}
m(\mu_{R}^{2})=m(\mu_{0}^{2})\left(\frac{\alpha_{s}(\mu_{R}^{2})}{\alpha_{s}(\mu_{0}^{2})}\right)^{\frac{\gamma_{m}^{(0)}}{2\beta_{0}}}
\end{eqnarray}
with the anomalous dimension of quark mass $\gamma_{m}^{(0)}=6C_{F}$ and the $\beta$ function $\beta_{0}=\frac{11 N_{c}-2 N_{f}}{3}$ as defined above, we can get the $\mu_{\pi}$ for pion and $\mu_{K}$ for kaon with the different energy scale $\mu_{R}^{2}=W^{2}$. The masses of current quarks are $m_{\mu} (1\ \textrm{GeV})=m_{d} (1\ \textrm{GeV})=4$ MeV and $m_{s} (1\ \textrm{GeV})=140$ MeV \cite{Ball3}. The masses of pseudoscalar mesons $m_{\pi}=139.6$ MeV and $m_{K}=493.7$ MeV are quoted from PDG \cite{Olive}. With the help of Eq.\ (\ref{3.2}) and Eq.\ (\ref{3.3}), we work out $2.44\ \textrm{GeV}\leq\mu_{\pi}\leq 3.66\ \textrm{GeV}$ and $1.69\ \textrm{GeV}\leq\mu_{K}\leq 2.55\ \textrm{GeV}$ in the energy scale $W\in(1,6)$.

\begin{figure}
\centering
     \includegraphics[width=0.48\textwidth, height=0.35\textwidth]{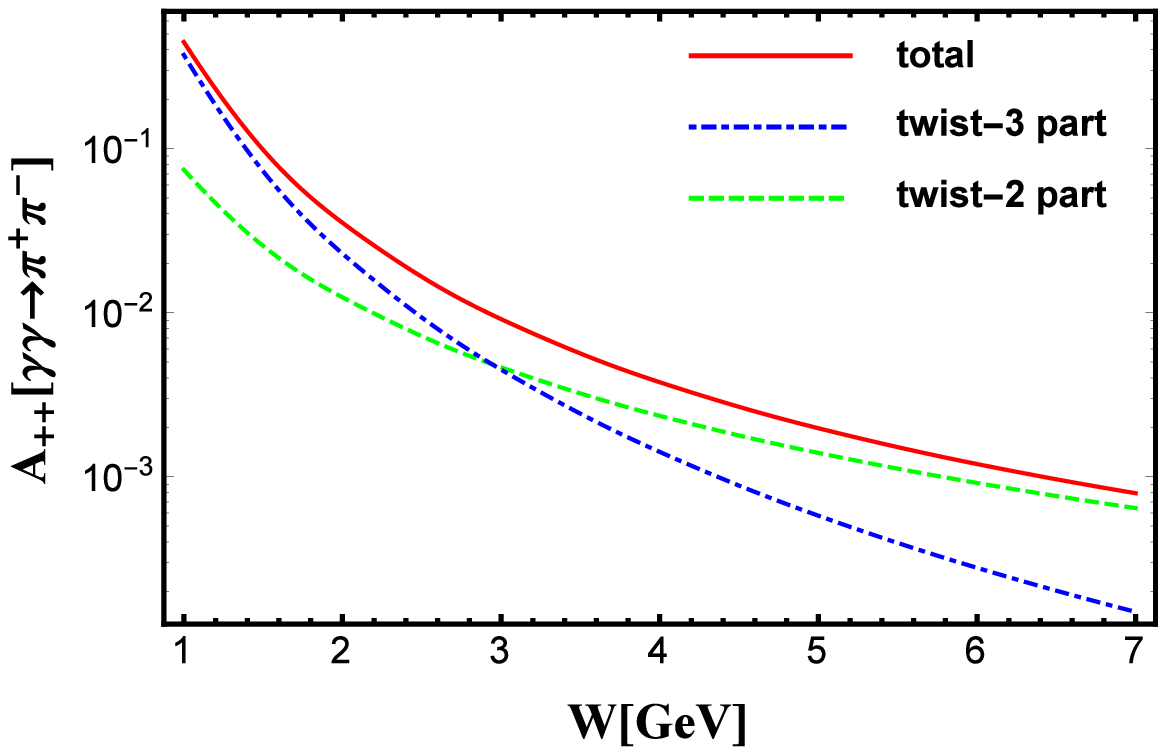}
     \includegraphics[width=0.48\textwidth, height=0.35\textwidth]{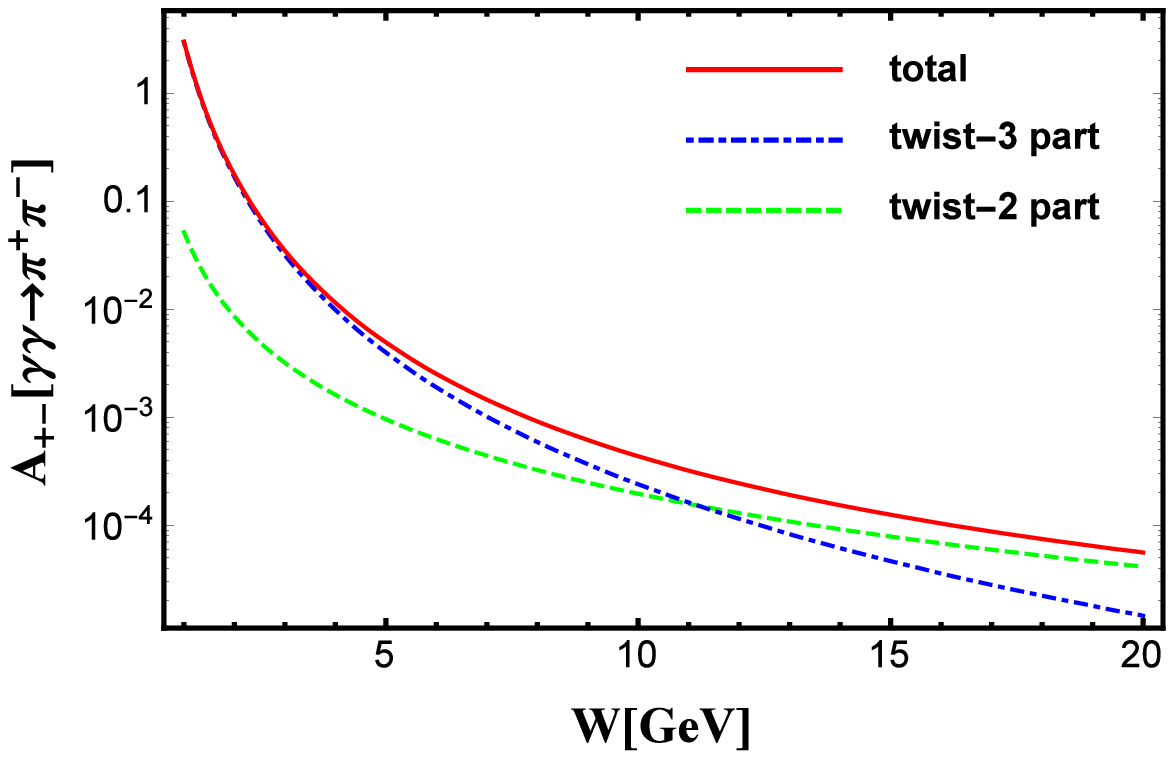}
     \includegraphics[width=0.48\textwidth, height=0.35\textwidth]{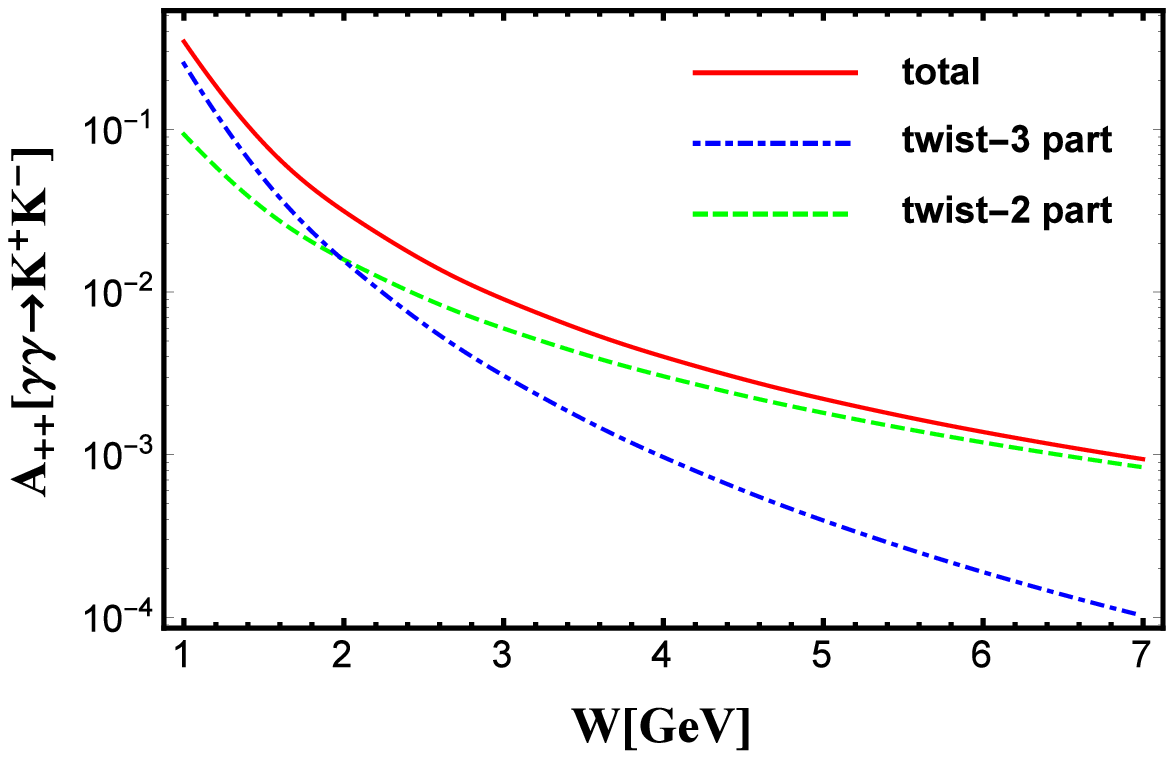}
     \includegraphics[width=0.48\textwidth, height=0.35\textwidth]{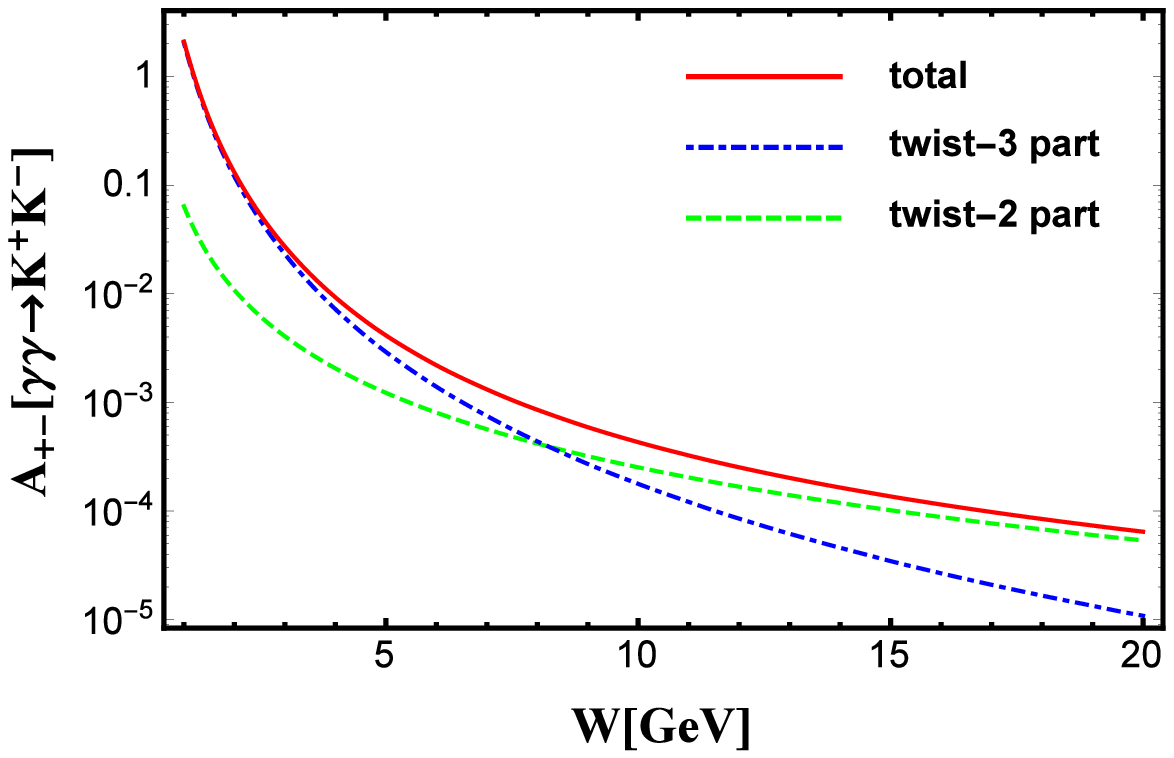}
\caption{Dependence of the prediction for $\gamma\gamma\rightarrow\overline{M}M$ transitional matrix element $A_{\lambda_{1}\lambda_{2}}$ on the energy $W$ with the scattering angle $cos\theta=0$ in the pion and kaon case, respectively.}
\label{TME}
\end{figure}

The transitional matrix elements $A_{\lambda_{1}\lambda_{2}}$ of the two-pion and two-kaon processes are shown in Fig. \ref{TME} with the two-photon energy $W$ as a variable parameter. Here we fix the scattering angle $cos\theta=0$ and choose $a_{n}^{i}$ as the central value in Eqs. (\ref{2.22}) for mesons distribution amplitudes and $\Lambda_{QCD}=0.2\ \textrm{GeV}$ for the strong coupling constant. Considering the polarization states of two photons, there are some relationships $A_{++}=A_{--}$ and $A_{+-}=A_{-+}$ in our calculation. In Fig.3, the left two figures and the right two figures are the transitional matrix elements $A_{++}$ and $A_{+-}$ for the pion and kaon case, respectively. The green dashed curves are the contributions from twist-2 distribution amplitudes, the blue dotdashed curves are contributions from twist-3 distribution amplitudes and the red solid curves are total contributions with twist-2 part and twist-3 part. Comparing with the leading-twist contribution, we can see that the twist-3 contribution is suppressed in the transitional matrix element $A_{++}$ as the energy $W>$ 3 GeV(2 GeV) for pion(kaon), while the similar condition is occurred in the transitional matrix element $A_{+-}$ as the energy $W>$ 11 GeV(8 GeV) for pion(kaon).

\begin{figure}
\centering
     \includegraphics[width=0.48\textwidth, height=0.35\textwidth]{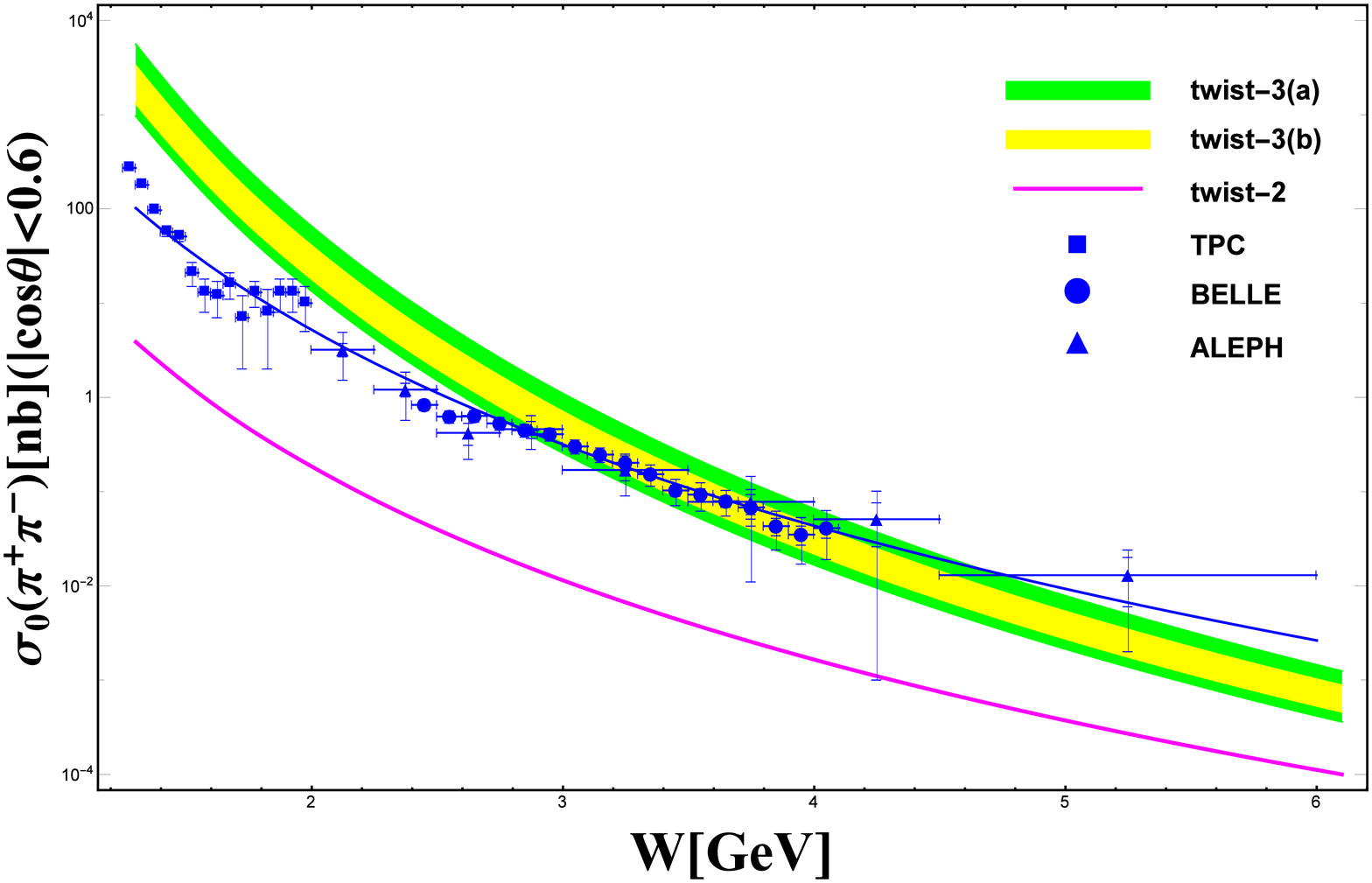}
     \includegraphics[width=0.48\textwidth, height=0.35\textwidth]{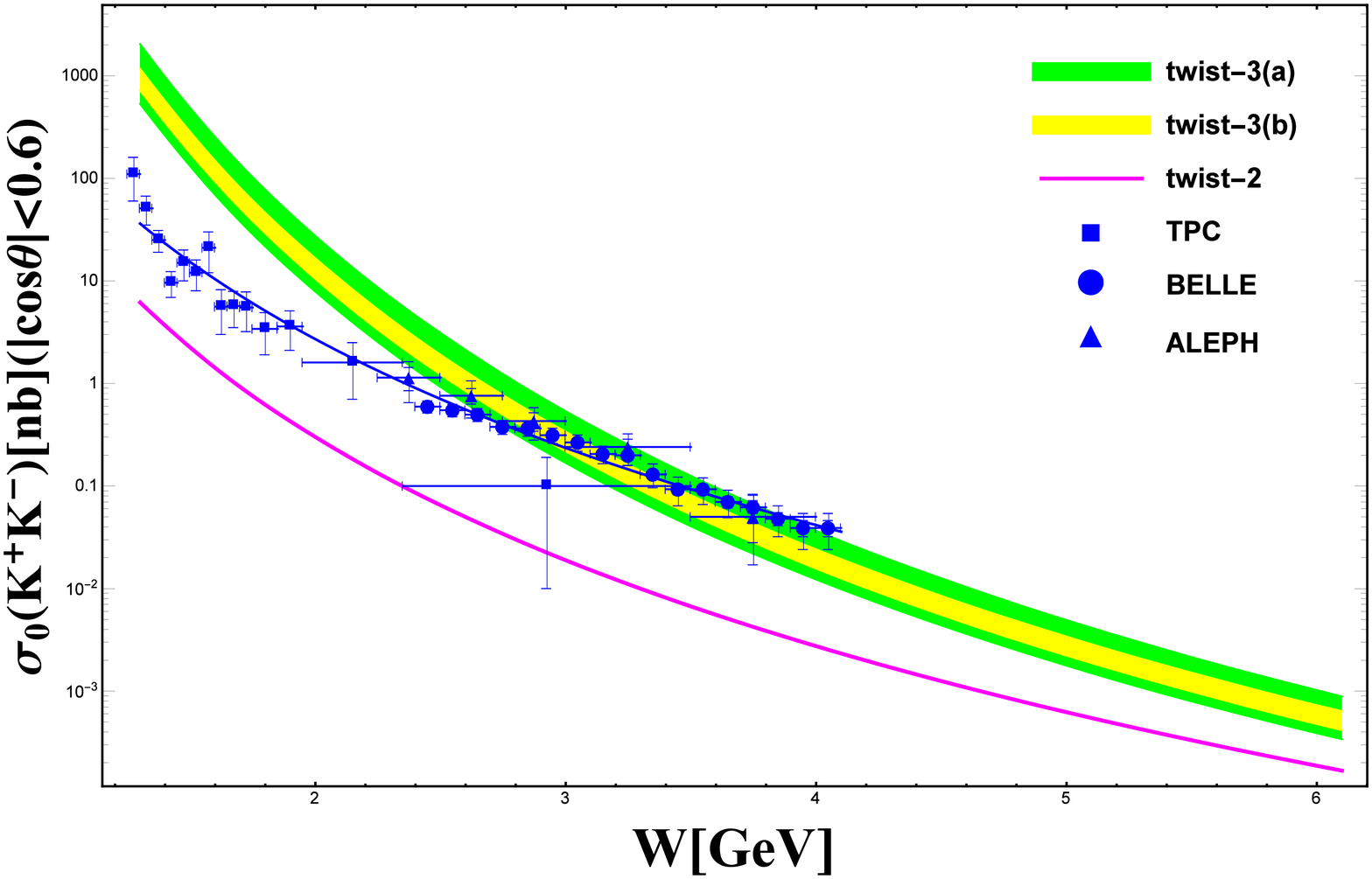}
\caption{Cross section for $\gamma\gamma\rightarrow\overline{M}M$ in the c.m. angular region $|cos\theta|<0.6$. The blue solid line is the result of fit for the data among relevant ranges.}
\label{cspik}
\end{figure}

Integrating over the scattering angular with $|\cos\theta|<0.6$, the cross sections $\sigma_{0}(\gamma\gamma\rightarrow\pi^{+}\pi^{-})$ and $\sigma_{0}(\gamma\gamma\rightarrow K^{+}K^{-})$ are shown in Fig. \ref{cspik}. The banded structures are the contribution from twist-3 parts. The green band is noted twist-3(a) with the variable $a_{n}^{i}$ and $\alpha_{s}$ with $\Lambda_{QCD}\in(0.15,0.3)$ GeV. The yellow band is noted twist-3(b) with the variable $a_{n}^{i}$ at $\Lambda_{QCD}=0.2\ \textrm{GeV}$ for $\alpha_{s}$. Here we can see that the strong coupling constant $\alpha_{s}$ has a little effect on the lower limit but it has an obviously effect to the upper limit from the areas of two bands. The choice of distribution amplitudes has a significant influence on the cross section from the yellow band. The magenta solid line named twist-2 stands for the contribution from twist-2 part with $\Lambda_{QCD}=0.2\ \textrm{GeV}$ for $\alpha_{s}$ and $a_{n}^{i}$ chosen as the central value in Eqs.(\ref{2.22}).
The experimental data from TPC \cite{Aihara}, BELLE \cite{Nakazawa} and ALEPH \cite{Heister} are also displayed simultaneously in Fig. \ref{cspik} and we find that the twist-3 correction to the cross section is markedly improved, even than an order of magnitude enhance to compare with leading-twist contribution and one-loop correction \cite{Duplancic}. Especially, our results of cross sections for the pion and kaon channels are in good agreement with the experimental data in the energy $W > 2.8$ GeV.

Our curves differ significantly from the predictions made before due to considering the contributions from twist-3 parts. We know the parametrization about the $W$ dependence of the cross section, which has the form of $\sigma(\gamma\gamma\rightarrow M^{+}M^{-})\propto W^{-n_{M}}$. In Ref. \cite{Nakazawa}, BELLE Collaboration announce that they find $n_{\pi}=7.9\pm0.4\pm1.5$ and $n_{K}=7.3\pm0.3\pm1.5$ for $3.0\ \textrm{GeV}<W<4.1\ \textrm{GeV}$. In Ref. \cite{Duplancic}, the NLO results give the power $n_{\pi}=n_{K}=6.9(7.4)$ for $\mu^{2}_{R}=W^{2}(W^{2}/15)$. On the other hand, we carry out a simple theoretic fitting for the experimental data, which are mentioned in Fig. \ref{cspik} among $1.25\ \textrm{GeV}<W<6(4.1)\ \textrm{GeV}$ for pion(kaon), and obtain $n_{\pi}=6.91$ and $n_{K}=6.02$ corresponding to the blue solid curves in Fig. \ref{cspik}. At the same time, we find that the powers are $n_{\pi}=6.76$ and $n_{K}=6.73$ from our twist-2 contributions and the powers are $n_{\pi}=9.63$ and $n_{K}=9.20$ from our twist-3 contributions. The analysis of the powers shows that our predictions from twist-3 parts change faster than the experimental data but their values are the similar magnitude with the data from BELLE, ALEPH, TPC, and vice versa for twist-2 parts.

\begin{figure}
\centering
     \includegraphics[width=0.6\textwidth, height=0.4\textwidth]{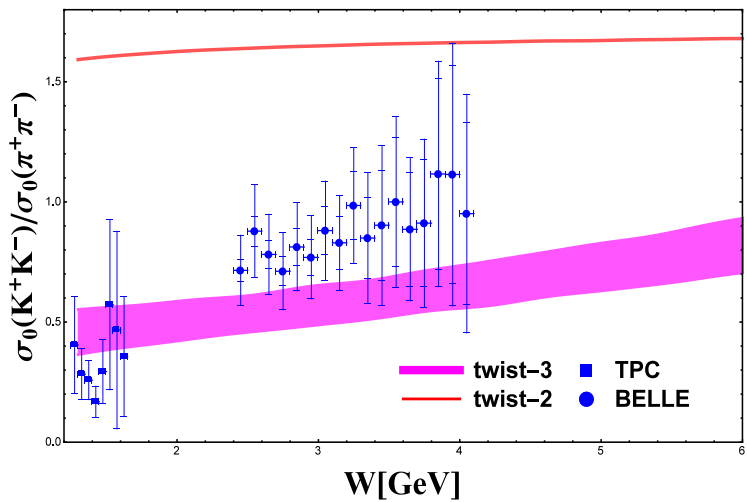}
\caption{Cross section ration of $\gamma\gamma\rightarrow\pi^{+}\pi^{-}$ to $\gamma\gamma\rightarrow K^{+}K^{-}$. The experimental points are from TPC\cite{Aihara} and BELLE\cite{Nakazawa}.}
\label{csratio}
\end{figure}

The ratio $K^{+}K^{-}$ to $\pi^{+}\pi^{-}$ is showed in Fig. \ref{csratio}. The experimental points are calculated from the data of TPC \cite{Aihara} with the energy $W\in (1.2, 2.0)$ GeV and BELLE \cite{Nakazawa} with the energy $W\in (2.4, 4.1)$ GeV. The red solid curve is the contribution of twist-2 distribution amplitude from the first three terms Gegenbauer polynomials with the $\Lambda_{QCD}=0.2$ and $a_{n}^{i}$ taken as the center value of Eq.(\ref{2.22}) and it is larger than the experimental data. The prediction in Ref. \cite{Benayoun} is approximately equals to $1.06$ and the one-loop prediction $f_{K}^{4}/f_{\pi}^{4}=2.23$ \cite{Duplancic} coincides with the BL estimate \cite{Brodsky2} with the asymptotic leading twist distribution amplitude, while the Belle measured value is $0.89\pm0.04\pm0.15$ in the energy $W\in (3.0, 4.1)$ GeV. The magenta band is from our twist-3 correction with the variable $a_{n}^{i}$ and $\alpha_{s}$ with $\Lambda_{QCD}\in(0.15,0.3)$ GeV and it is in agreement with the experimental data.

\begin{figure}
\centering
     \includegraphics[width=0.45\textwidth, height=0.45\textwidth]{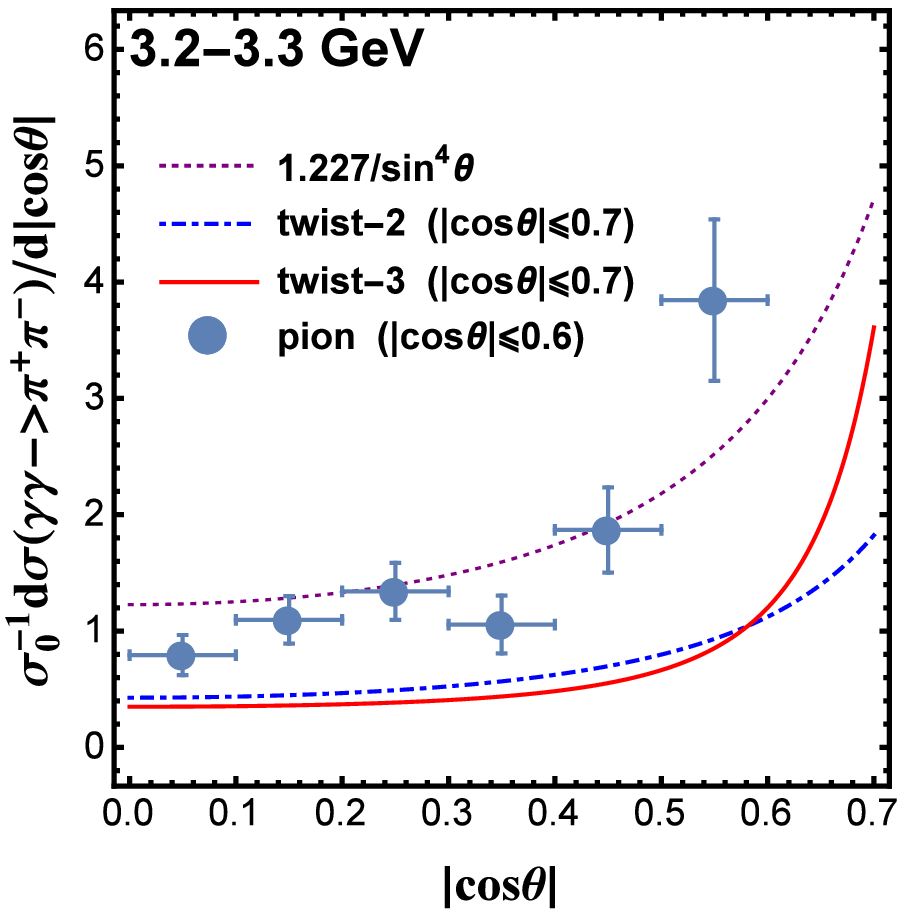}
     \includegraphics[width=0.45\textwidth, height=0.45\textwidth]{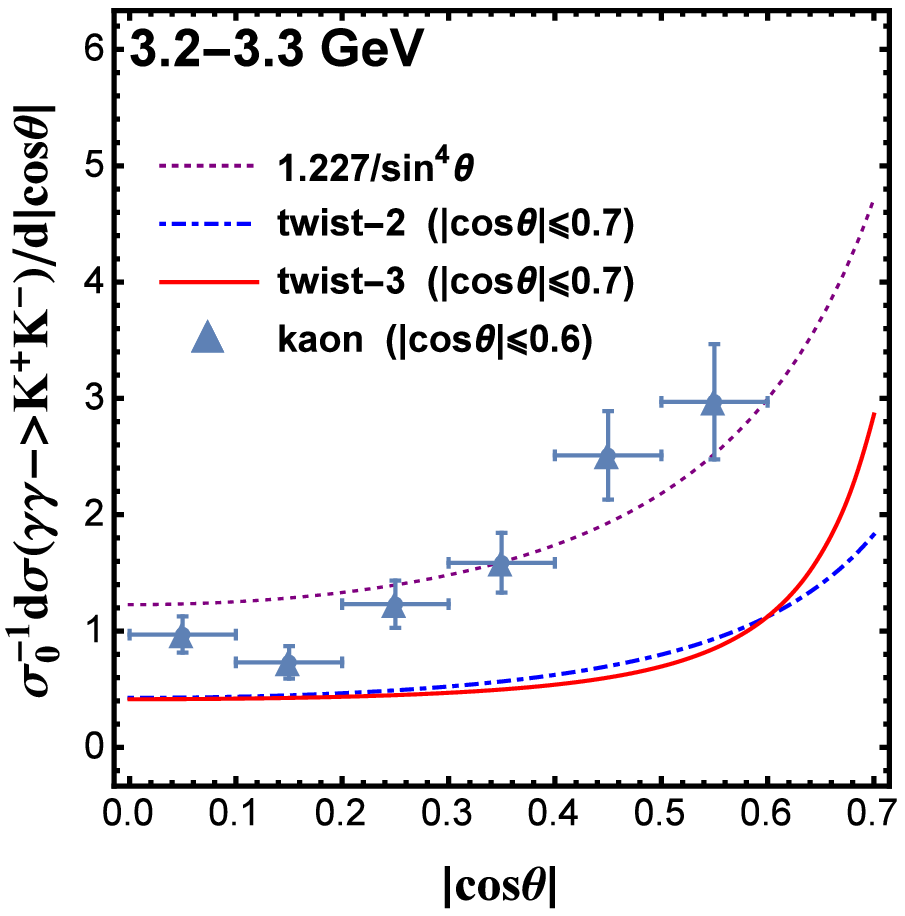}
\caption{Angular dependence of the cross section, $\sigma_{0}^{-1} d\sigma/ d|\cos\theta|$ for $\pi^{+}\pi^{-},K^{+}K^{-}$. The experimental points are from Belle Collaboration\cite{Nakazawa}.}
\label{ADpik}
\end{figure}

In Fig. \ref{ADpik}, we display the angular dependence of cross section, $\sigma_{0}^{-1}d\sigma/d|\cos\theta|$ for the $\pi^{+}\pi^{-}$ and $K^{+}K^{-}$ processes, respectively. There is no obvious change to our prediction for the ratio by varying $W$ from $2.4$ to $4.1$, where the Belle data \cite{Nakazawa}  are covered. To simplify our analysis, we only discuss our prediction and the Belle data at the $W=3.2\sim3.3$ GeV region. The discrete points come from the experimental data in the angular region $|\cos\theta|\leq0.6$. The dotted curves indicate the expectation from a $\sin^{-4}\theta$ behavior predicted by Brodsky and Lepage \cite{Brodsky2}. The dot-dashed and solid curves correspond to the twist-2 and twist-3 contribution in the angular $|\cos\theta|\leq0.7$, respectively. One in particular is to expand the scattering angle to $|\cos\theta|\leq0.7$ in our drawing and it is aimed at reflecting the distinction between twist-2 and twist-3 parts at the large scattering angle. Since the $\sigma_{0}^{-1} d\sigma/ d|\cos\theta|$ for $|\cos\theta|\leq0.7$ is less than $5\%$ for $|\cos\theta|\leq0.6$ and it has almost no influence for further analysis. We can see that our results are consistent with the experimental data and the curves have the similar changes with $\sin^{-4}\theta$ in the small angular area. These ratios are independent for the strong coupling constant and it is very difficult to distinguish them by varying the distribution amplitudes in our calculation. It is worth noting that the $\sigma_{0}^{-1}d\sigma/d|\cos\theta|$ with one-loop correction \cite{Nizic,Duplancic} is in very good agreement with the data.

\section{conclusion}
In this work, we recalculate the two-photon annihilation into two pseudoscalar mesons processes with the corrections of the two-particle twist-3 distribution amplitudes of pseudoscalar mesons in the standard hard-scattering approach. In order to avoid the end-point singularity from twist-3 distribution amplitudes, we take distribution amplitudes of meson with BHL prescription. The twist-3 corrections of cross sections for $\gamma\gamma\rightarrow\pi^{+}\pi^{-}, K^{+}K^{-}$ are markedly improved, even than an order of magnitude enhance to compare with the twist-2 contributions. The results show that the contributions from twist-3 parts are actually not power suppressed comparing with the leading-twist contributions, because momentum transfer is not enough large in those processes. While the cross sections with twist-3 corrections have the similar changes and are at the same order in magnitude with the data by varying the center-of-mass energy $W$ from $1$ GeV to $6$ GeV. We also discuss the cross section ratio $\sigma_{0}(K^{+}K^{-})/\sigma_{0}(\pi^{+}\pi^{-})$ and find it close to experimental results. Numerical analysis for the angular dependence of $\sigma_{0}^{-1} d\sigma/ d|\cos\theta|$ shows that the ratios are the independence of distribution amplitudes and are a good agreement with the experimental data in the small-angle area, but the ratios with the twist-3 corrections increase faster than the ones from the twist-2 parts in the large-angle area.

\section{acknowledgements}
We gratefully thanks to F.G. Cao, H.Q. Zhou, W.L. Sang and L. Cao for many helpful discussions. This work was supported by the Natural Science Foundation of China, Grant Number 11005087, 11175146, the Fundamental Research Funds for the Central Universities, Grant Number XDJK2014C169, XDJK2016C067, the Open Project Program of State Key Laboratory of Theoretical Physics, Grant Number Y4KF081CJ1.

\newpage
\textbf{Appendix A: The expression of invariant amplitudes $T^{ij}_{\lambda_{1}\lambda_{2}}(x,y,W,\theta)$.}

In this appendix, we present the detailed formulas of invariant amplitudes $T^{ij}_{\lambda_{1}\lambda_{2}}(x,y,W,\theta)$ with the contributions from the two-particle twist-2 and twist-3 distribution amplitudes in the $\gamma\gamma\rightarrow\pi^{+}\pi^{-}$ process and the relevant expressions for the $\gamma\gamma\rightarrow K^{+}K^{-}$ process can be obtained by making the replacements of $f_{\pi}\rightarrow f_{K}$, $\mu_{\pi}\rightarrow \mu_{K}$ and $e_{d}\rightarrow e_{s}$ in the following formulas.

The leading-twist hard-scattering amplitudes for the $\pi^{+}\pi^{-}$ channel are expressed as
\begin{equation*}
   \begin{split}
     T^{\pi\pi}_{++}=T^{\pi\pi}_{--}=\frac{16\pi^{2}\alpha\alpha_{s}(\mu_{R}^{2})C_{F}f_{\pi}^{2}[-y+x(-1+2y)](e_{d}-e_{u})^{2}}{3(-1+t^{2})W^{2}(-1+x)x(-1+y)y},
   \end{split}
\end{equation*}
\begin{equation*}
    \begin{split}
      T^{\pi\pi}_{+-}=&T^{\pi\pi}_{-+}=\frac{4\pi^{2}\alpha\alpha_{s}(\mu_{R}^{2})C_{F}f_{\pi}^{2}}{3(-1+t^{2})W^{2}(-1+x)x(-1+y)y}
      \bigg{\{}2[-1+x+y-4xy+t^{2}(-1+x+y)]e_{d}^{2}\\
      +&\frac{1}{[x(1+t-2y)+y-ty][-(1+t)y+x(-1+t+2y)]}\Big{\{}4\big{\{}-(-1+t^{2})(-1+y)y^{2}\\
      +&x^{3}(-1+2y)(-1+t^{2}+8y-8y^{2})+x^{2}[-1+11y-32y^{2}+24y^{3}+t^{2}(1+5y-8y^{2})]\\
      +&xy[-2+11y-10y^{2}+t^{2}(-6+5y+2y^{2})]\big{\}}e_{d}e_{u}\Big{\}}\\
      -&[6-6y+2t^{2}(-1+x+y)+x(-6+8y)]e_{u}^{2}
      \bigg{\}},
     \end{split}
\end{equation*}
where $\alpha$ and $\alpha_{s}$ mean the electromagnetic coupling constant and the strong coupling constant, respectively. The variables $x$ and $y$ are the momentum fractions from the final pseudoscalar mesons. The quark charges are $e_{u}=\frac{2}{3}$ for the $u$ quark and $e_{d(s)}=-\frac{1}{3}$ for the $d(s)$ quark. The color factor is $C_{F}=\frac{N_{C}^{2}-1}{2N_{C}}$ with the color number $N_{C}=3$. If we make a replacement of $t\rightarrow\cos\theta$ in the above expression, we reduce them to the new formulas that have the same forms with the theoretical prediction of V.L. Chernyak \cite{Chernyak4,Chernyak7}, S.J. Brodsky \cite{Brodsky2} and Bene Ni$\check{\textrm{z}}$i$\acute{\textrm{c}}$ \cite{Nizic} on the leading-twist order.

It is convenient to take $\cos\theta$ as $t$ in the numerical analysis of the twist-3 parts and the twist-3 hard-scattering amplitudes are depicted as follows

\begin{equation*}
    \begin{split}
      T^{pp}_{++}=&T^{pp}_{--}=-\frac{8\pi^{2}\alpha\alpha_{s}(\mu_{R}^{2})C_{F}f_{\pi}^{2}\mu_{\pi}^{2}}{3(-1+t^{2})W^{4}(-1+x)x(-1+y)y}
      \bigg{\{}-2\{2y+x[2+(-3+t^{2})y]\}e_{d}^{2}\\
      +&\frac{1}{[x(1+t-2y)+y-ty][-(1+t)y+x(-1+t+2y)]}\Big{\{}8\big{\{}(-1+t^{2})(-1+y)y^{2}\\
      +&x^{3}(-1+2y)[1-y+y^{2}+t^{2}(-1-y+y^{2})]+xy[2-4y+3y^{2}-t^{2}(-2+2y+y^{2})]\\
      -&x^{2}[-1+4y-4y^{2}+3y^{3}+t^{2}(1+2y-8y^{2}+3y^{3})]\big{\}}e_{d}e_{u}\Big{\}}\\
      -&2[1+x+t^{2}(-1+x)(-1+y)+y-3xy]e_{u}^{2}
      \bigg{\}},
     \end{split}
\end{equation*}
\begin{equation*}
    \begin{split}
      T^{pp}_{+-}=&T^{pp}_{-+}=\frac{8\pi^{2}\alpha\alpha_{s}(\mu_{R}^{2})C_{F}f_{\pi}^{2}\mu_{\pi}^{2}}{3(1-t^{2})W^{4}}
      \Big{\{}\frac{2[-1+x(2-3y)+2y+t^{2}(-1+xy)]e_{d}^{2}}{(-1+x)^{2}(-1+y)^{2}}\\
      -&\frac{8(1+t^{2})[-y+x(-1+2y)]e_{d}e_{u}}{[x(1+t-2y)+y-ty][-(1+t)y+x(-1+t+2y)]}\\
      +&\frac{2\{x+[1+t^{2}(-1+y)-3y]+y-t^{2}y\}e_{u}^{2}}{x^{2}y^{2}}
      \Big{\}},
     \end{split}
\end{equation*}

\begin{equation*}
    \begin{split}
      T^{p\sigma}_{++}=&T^{p\sigma}_{--}=-\frac{16\pi^{2}\alpha\alpha_{s}(\mu_{R}^{2})C_{F}f_{\pi}^{2}\mu_{\pi}^{2}}{9(-1+t^{2})^{2}W^{4}(-1+x)^{2}x^{2}}
      \bigg{\{}-\frac{(-1+t^{2})(-1+x)(1+t^{2}x)e_{d}^{2}}{-1+y}\\
      +&\frac{1}{(-1+y)y[-x(1+t-2y)+(-1+t)y](x-tx+y+ty-2xy)}\\
      &\Big{\{}2\big{\{}(-1+t^{2})(-1+y)y^{2}+xy[-1+y+2t^{2}y-2y^{2}+t^{4}(1-3y+2y^{2})]\\
      +&x^{5}[1-4y+6y^{2}-4y^{3}+t^{4}(-1+2y)-6t^{2}y(1-3y+2y^{2})]\\
      +&x^{2}y[-3+8y+2y^{2}+t^{4}(1+10y-12y^{2})+2t^{2}(1-13y+9y^{2})]\\
      +&x^{3}[-1+6y-6y^{2}-8y^{3}+t^{2}(2-6y+42y^{2}-32y^{3})+t^{4}(-1-8y+4y^{2}+8y^{3})]\\
      +&x^{4}[t^{4}(2+3y-8y^{2})+y(1-6y+10y^{2})+2t^{2}(-1+6y-21y^{2}+15y^{3})]\big{\}}e_{d}e_{u}\Big{\}}\\
      +&\frac{(-1+t^{2})[-1+t^{2}(-1+x)]xe_{u}^{2}}{y}
      \bigg{\}},
     \end{split}
\end{equation*}
\begin{equation*}
    \begin{split}
      T^{p\sigma}_{+-}=&T^{p\sigma}_{-+}=-\frac{8\pi^{2}\alpha\alpha_{s}(\mu_{R}^{2})C_{F}f_{\pi}^{2}\mu_{\pi}^{2}}{9(-1+t^{2})^{2}W^{4}(-1+x)^{3}x^{3}}
      \bigg{\{}\frac{(-1+t^{2})x^{2}[-1+2x+t^{2}(-1-2x+2x^{2})]e_{d}^{2}}{-1+y}\\
      +&\frac{1}{(-1+y)y[x(1+t-2y)+y-ty][-(1+t)y+x(-1+t+2y)]}\\
      &\Big{\{}4(-1+x)^{2}x^{2}\big{\{}-(-1+t^{2})y^{2}[-1+2y+t^{2}(-3+4y)]\\
      +&x^{3}[1-4y+6y^{2}-4y^{3}+t^{4}(-1+2y)-6t^{2}y(1-3y+2y^{2})]\\
      +&xy[-1+4y-6y^{2}+t^{2}(2+8y-10y^{2})+t^{4}(-1-4y+8y^{2})]\\
      +&x^{2}y[1+t^{4}(5-8y)-4y+6y^{2}+2t^{2}(1-10y+9y^{2})]\big{\}}e_{d}e_{u}\Big{\}}\\
      +&\frac{(-1+t^{2})(-1+x)^{2}[1-2x+t^{2}(-1-2x+2x^{2})]e_{u}^{2}}{y}
      \bigg{\}},
     \end{split}
\end{equation*}

\begin{equation*}
    \begin{split}
      T^{\sigma p}_{++}=&T^{\sigma p}_{--}=-\frac{16\pi^{2}\alpha\alpha_{s}(\mu_{R}^{2})C_{F}f_{\pi}^{2}\mu_{\pi}^{2}}{9(-1+t^{2})^{2}W^{4}(-1+y)^{2}y^{2}}
      \bigg{\{}-\frac{(-1+t^{2})(-1+y)(1+t^{2}y)e_{d}^{2}}{-1+x}\\
      +&\frac{1}{(-1+x)x[x(1+t-2y)+y-ty][-(1+t)y+x(-1+t+2y)]}\\
      &\Big{\{}2\big{\{}-(-1+t^{2})(-1+y)y^{3}[1+t^{2}(-1+y)+y]+xy[-1-3y+6y^{2}+y^{3}-4y^{4}\\
      +&2t^{2}y(1-3y+6y^{2}-3y^{3})+t^{4}(1+y-8y^{2}+3y^{3}+2y^{4})]-x^{3}(-1+2y)\\
      &[1+2y^{2}-4y^{3}+2y^{4}+t^{4}(1+4y-4y^{2})+2t^{2}(-1-2y+5y^{2}-6y^{3}+3y^{4})]\\
      +&x^{2}[-1+y+8y^{2}-6y^{3}-6y^{4}+6y^{5}+t^{4}(-1-3y+10y^{2}+4y^{3}-8y^{4})\\
      +&2t^{2}(1+y-13y^{2}+21y^{3}-21y^{4}+9y^{5})]\big{\}}e_{d}e_{u}\Big{\}}\\
      +&\frac{(-1+t^{2})[-1+t^{2}(-1+y)]ye_{u}^{2}}{x}
      \bigg{\}},
     \end{split}
\end{equation*}
\begin{equation*}
    \begin{split}
      T^{\sigma p}_{+-}=&T^{\sigma p}_{-+}=-\frac{8\pi^{2}\alpha\alpha_{s}(\mu_{R}^{2})C_{F}f_{\pi}^{2}\mu_{\pi}^{2}}{9(-1+t^{2})^{2}W^{4}(-1+y)^{3}y^{3}}
      \bigg{\{}\frac{(-1+t^{2})y^{2}[-1+2y+t^{2}(-1-2y+2y^{2})]e_{d}^{2}}{-1+x}\\
      +&\frac{1}{(-1+x)x[x(1+t-2y)+y-ty][-(1+t)y+x(-1+t+2y)]}\\
      &\Big{\{}4(-1+y)^{2}y^{2}\big{\{}-(-1+t^{4})y^{3}+xy[-1+y-4y^{2}+t^{2}(2+2y-6y^{2})\\
      +&t^{4}(-1+5y+2y^{2})]+2x^{3}[1-3y+3y^{2}-2y^{3}+t^{4}(-2+4y)+t^{2}(1-5y+9y^{2}-6y^{3})]\\
      +&x^{2}[-1+4y-4y^{2}+6y^{3}+t^{4}(3-4y-8y^{2})+2t^{2}(-1+4y-10y^{2}+9y^{3})]\big{\}}e_{d}e_{u}\Big{\}}\\
      +&\frac{(-1+t^{2})(-1+y)^{2}[1-2y+t^{2}(-1-2y+2y^{2})]ye_{u}^{2}}{x}
      \bigg{\}},
     \end{split}
\end{equation*}

\begin{equation*}
    \begin{split}
    T^{\sigma\sigma}_{++}=&T^{\sigma\sigma}_{--}=\frac{2\pi^{2}\alpha\alpha_{s}(\mu_{R}^{2})C_{F}f_{\pi}^{2}\mu_{\pi}^{2}}{27(1-t^{2})^{2}W^{4}}
    \Bigg{\{}\frac{2(1+t^{2})[x(-1+y)-y]e_{d}^{2}}{(-1+x)^{2}x(-1+y)^{2}y}+(-1+t^{2})^{2}\\
    &\bigg{\{}-\Big{\{}\frac{1}{(1+t)^{4}x^{2}(-1+y)^{2}[x(1+t-2y)+y-ty]^{2}}\\
    &\big{\{}4\{-(-1+t)[-1+t(-1+y)-y](-1+y)y+x^{3}[-1+4y-6y^{2}+4y^{3}\\
    +&t^{2}(-1+2y)+t(-2+6y-6y^{2})]+x[2-2y-2y^{2}+4y^{3}-t^{3}(-1+y^{2})\\
    +&t(5-4y+3y^{2}-6y^{3})+2t^{2}(2-y-2y^{2}+y^{3})]+x^{2}[-1+t^{3}(-1+y)\\
    +&4y^{2}-6y^{3}+t^{2}(-3+6y-4y^{2})+t(-3+5y-4y^{2}+6y^{3})]\}\big{\}}\Big{\}}\\
    +&\frac{1}{(-1+t)^{4}x^{2}(-1+y)^{2}(x-tx+y+ty-2xy)^{2}}
    \big{\{}4\{(1+t)(-1+y)y[1+t(-1+y)+y]\\
    +&x^{3}[1+t^{2}(1-2y)-4y+6y^{2}-4y^{3}+t(-2+6y-6y^{2})]+x[-t^{3}(-1+y^{2})\\
    +&t(5-4y+3y^{2}-6y^{3})+2(-1+y+y^{2}-2y^{3})-2t^{2}(2-y-2y^{2}+y^{3})]\\
    +&x^{2}[1+t^{3}(-1+y)-4y^{2}+6y^{3}+t^{2}(3-6y+4y^{2})+t(-3+5y-4y^{2}+6y^{3})]\}\big{\}}\\
    -&\frac{1}{(-1+t)^{4}(-1+x)^{2}y^{2}[x(1+t-2y)+y-ty]^{2}}\big{\{}4\{(-1+t)^{2}(-2+t-y)(-1+y)y\\
    +&x^{3}[-1+4y-6y^{2}+4y^{3}+t^{2}(-1+2y)+t(-2+6y-6y^{2})]\\
    -&(-1+t)x[1-2y+t^{2}y^{2}+4y^{3}+t(1+2y-5y^{2}-2y^{3})]\\
    +&x^{2}[t^{3}y+t^{2}(2-4y-4y^{2})-2y(1-2y+3y^{2})+t(2-3y+4y^{2}+6y^{3})]\}\big{\}}\\
    +&\frac{1}{(1+t)^{4}(-1+x)^{2}y^{2}(x-tx+y+ty-2xy)^{2}}\big{\{}4\{(1+t)^{2}(-1+y)y(2+t+y)\\
    +&x^{3}[1+t^{2}(1-2y)-4y+6y^{2}-4y^{3}+t(-2+6y-6y^{2})]\\
    -&(1+t)x[1-2y+t^{2}y^{2}+4y^{3}+t(-1-2y+5y^{2}+2y^{3})]\\
    +&x^{2}[t^{3}y+2y(1-2y+3y^{2})+t^{2}(-2+4y+4y^{2})+t(2-3y+4y^{2}+6y^{3})]\}\big{\}}\bigg{\}}e_{d}e_{u}\\
    +&\frac{2(1+t^{2})(-1+xy)e_{u}^{2}}{(-1+x)x^{2}(-1+y)y^{2}}
    \Bigg{\}},
    \end{split}
\end{equation*}

\begin{equation*}
    \begin{split}
    T^{\sigma\sigma}_{+-}=&T^{\sigma\sigma}_{-+}=\frac{4\pi^{2}\alpha\alpha_{s}(\mu_{R}^{2})C_{F}f_{\pi}^{2}\mu_{\pi}^{2}}{27(-1+t^{2})W^{4}}
    \Bigg{\{}\frac{(1+t^{2})[2+x(-1+y)-y]e_{d}^{2}}{(-1+x)^{2}x(-1+y)^{2}y}-\frac{1}{(-1+t^{2})^{2}}\\
    &2\Big{\{}\frac{1}{x^{2}(-1+y)^{2}[x(-1+t-2y)+y-ty]^{2}}(-1+t)^{4}\{1(-1+t)[-3+t(-1+y)-y]y^{2}\\
    +&x^{3}[-1+4y-6y^{2}+4y^{3}+t^{2}(-1+2y)+t(-2+6y-6y^{2})]+x^{2}[-2+t^{3}(-1+y)\\
    +&6y-4y^{2}-6y^{3}+t^{2}(-4+6y-4y^{2})+t(-5+11y-4y^{2}+6y^{3})]+x[1-4y\\
    +&8y^{2}+4y^{3}-t^{3}(-1+y^{2})+t^{2}(3-4y^{2}+2y^{3})-t(-3+4y+3y^{2}+6y^{3})]\}\\
    -&\frac{1}{x^{2}(-1+y)^{2}(x-tx+y+ty-2xy)^{2}}(1+t)^{4}\{(1+t)y^{2}[3+t(-1+y)+y]\\
    +&x^{3}[1+t^{2}(1-2y)-4y+6y^{2}-4y^{3}+t(-2+6y-6y^{2})]+x^{2}[2+t^{3}(-1+y)\\
    -&6y+4y^{2}+6y^{3}+t^{2}(4-6y+4y^{2})+t(-5+11y-4y^{2}+6y^{3})]-x[1-4y\\
    +&8y^{2}+4y^{3}+t^{3}(-1+y^{2})+t^{2}(3-4y^{2}+2y^{3})+t(-3+4y+3y^{2}+6y^{3})]\}\\
    +&\frac{1}{(-1+x)^{2}y^{2}[x(1+t-2y)+y-ty]^{2}}(1+t)^{4}\{(-1+t)^{2}y[1+t(-1+y)-2y-y^{2}]\\
    -&(-1+t)xy[-4+(6-5t+t^{2})y-2(-2+t)y^{2}]\\
    +&x^{3}[-1+4y-6y^{2}+4y^{3}+t^{2}(-1+2y)+t(-2+6y-6y^{2})]\\
    +&x^{2}[-3+8y+t^{3}y-4y^{2}-6y^{3}+t^{2}(1-4y-4y^{2})+t(-2+3y+4y^{2}+6y^{3})]\}\\
    -&\frac{1}{(-1+x)^{2}y^{2}(x-tx+y+ty-2xy)^{2}}(-1+t)^{4}\{(1+t)^{2}y[-1+t(-1+y)+2y+y^{2}]\\
    -&(1+t)xy[-4+(6-5t+t^{2})y+2(2+t)y^{2}]\\
    +&x^{3}[1+t^{2}(1-2y)-4y+6y^{2}-4y^{3}+t(-2+6y-6y^{2})]\\
    +&x^{2}[3-8y+t^{3}y+4y^{2}+6y^{3}+t^{2}(-1+4y+4y^{2})+t(-2+3y+4y^{2}+6y^{3})]\}\Big{\}}e_{d}e_{u}\\
    +&\frac{(1+t^{2})(1+xy)e_{u}^{2}}{(-1+x)x^{2}(-1+y)y^{2}}
    \Bigg{\}}.
    \end{split}
\end{equation*}

\end{document}